\documentclass[journal=jpclcd]{achemso}
\setkeys{acs}{maxauthors=10,etalmode=truncate} 
\usepackage{amsmath}
\usepackage{color}
\usepackage{txfonts}
\usepackage{mathrsfs}
\usepackage{dcolumn}
\usepackage{bm}
\usepackage{multirow}
\usepackage{graphicx}
\usepackage{rotating}
\usepackage{layout}
\usepackage[version=3]{mhchem}
\usepackage{braket}
\usepackage{longtable}
\usepackage{setspace}

\newcommand{\beginsupplement}{%
        \setcounter{table}{0}
        \renewcommand{\thetable}{S\arabic{table}}%
        \setcounter{figure}{0}
        \renewcommand{\thefigure}{S\arabic{figure}}%
     }

\allowdisplaybreaks
\SectionNumbersOn
\usepackage{xfrac}

\usepackage[colorlinks,citecolor=blue,urlcolor=blue,bookmarks=false,hypertexnames=true]{hyperref} 

\title{Reaching high accuracy for energetic properties at second-order perturbation cost by merging self-consistency and spin-opposite scaling}
\date{\today}
\author{Nhan Tri Tran}
\affiliation{University of Science, Vietnam National University, Ho Chi Minh City, Vietnam}

\author{Hoang Thanh Nguyen}
\affiliation{Institute of Applied Mechanics and Informatics, Vietnam Academy of Science and Technology, Ho Chi Minh City, Vietnam}

\author{Lan Nguyen Tran}
\email{tnlan@hcmiu.edu.vn}
\affiliation{Department of Physics, International University, Ho Chi Minh City, Vietnam}
\affiliation{Vietnam National University, Ho Chi Minh City, Vietnam}

\begin{document}

\begin{abstract}
Quantum chemical methods dealing with challenging systems while retaining low computational costs have attracted attention. In particular, many efforts have been devoted to developing new methods based on the second-order perturbation that may be the simplest correlated method beyond Hartree-Fock. We have recently developed a self-consistent perturbation theory named one-body M{\o}ller-Plesset second-order perturbation theory (OBMP2) and shown that it can resolve issues caused by the non-iterative nature of standard perturbation theory. In the present work, we extend the method by introducing the spin-opposite scaling to the double-excitation amplitudes, resulting in the O2BMP2 method. We assess the O2BMP2 performance on the triple-bond N$_2$ dissociation, singlet-triplet gaps, and ionization potentials. O2BMP2 performs much better than standard MP2 and reaches the accuracy of coupled-cluster methods in all cases considered in this work.  

\begin{tocentry}
\includegraphics[width=5.0cm]{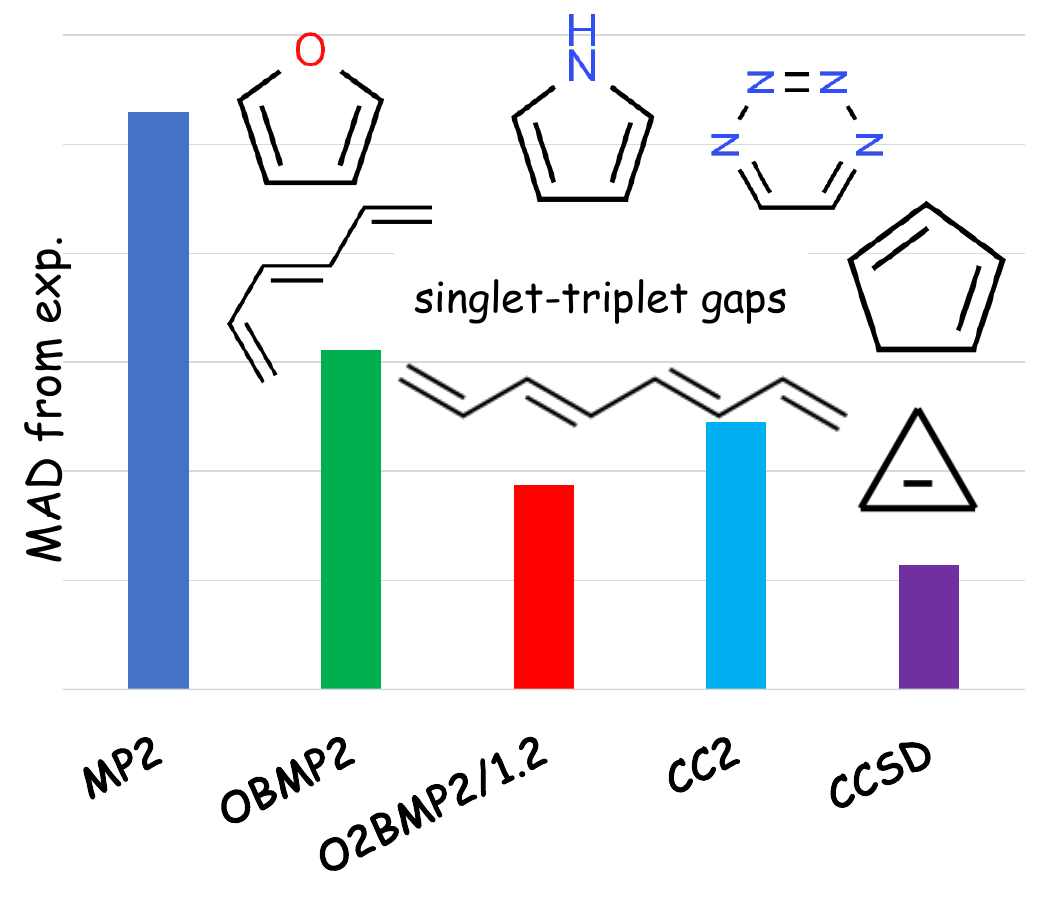}
\end{tocentry}

\end{abstract}

\maketitle

Second-order M{\o}ller-Plesset perturbation theory (MP2) on Hartree-Fock (HF) orbitals\cite{MP2} is the simplest correlated wave-function method. Its accuracy depends on the quality of reference wave functions, in particular, for open-shell systems\cite{MP2-JPCA2001, MP2-JCP2011}. To bypass the issue of poor references, many research groups have actively developed orbital-optimized MP2 (OOMP2) and its spin-scaled variants\cite{OOMP2-HeadGordon,OOMP2-Neese,OOMP2-Sherrill,Bozkaya2013-GradOMP2,Bozkaya2014-OMP2,Bozkaya2014-GradOMP2}. In these methods, orbitals are optimized by minimizing the Hylleraas functional. OOMP2 and its variants have outperformed standard MP2 calculations for numerous properties. Apart from wave-function methods, double-hybrid density functional (DHF) theory, in which a scaled perturbative correction is performed on top of hybrid density functional calculations, has attracted significant attention. These functionals are considered the fifth rung of the DFT Jacob's ladder and have been shown to outperform conventional functionals in many cases\cite{DHFs-WIRE2014,DHFs-IJC2020}. 

It is well-known that perturbation theory is inadequate for multi-reference systems, and the perturbative correlation energy diverges due to small gaps of orbital energies. To eliminate these issues, several regularization schemes that modify the MP2 amplitude with a function damping any divergent or excessively large correlations have been recently developed \cite{OOMP2-JCP2013,OOMP2-Molphys2017,OOMP2-JCTC2018,OOMP2-JPCL2021,OOMP2-JCTC2022}. It has been shown that regularized (orbital-optimized) MP2\cite{RMP2-JCP2022,RDHF-JPCL2022,DHFs-JPCL2023} can outperform standard MP2 across relevant chemical problems. In the meantime, numerous efforts are devoted to developing alternative approaches to resolving the abovementioned issues. These methods include Brillouin-Wigner perturbation theory (BWPT) and its size-consistent variant \cite{BWPT-JCP2023,RBWPT-2023}, retaining the excitation degree MP2 (REMP2) and its orbital-optimized variant \cite{REMP-JCP2019, OOREMP-JCTC2021}. Empirical spin-scaled methods, such as spin-component scaling (SCS) and spin-opposite scaling (SOS), have also been widely used to improve the performance of perturbation theory \cite{SCSMP2-2012}. Noticeably, SOS-MP2 does not only often improve the accuracy of MP2, but it is also less costly ($N^4$) than standard MP2 ($N^5$ ). 

In general, developing new methods based on low-cost perturbation theory able to deal with challenging systems is still highly desirable. Recently, we have developed a new self-consistent perturbation theory named one-body MP2 (OBMP2)\cite{OBMP2-JCP2013,OBMP2-JPCA2021, OBMP2-PCCP2022, OBMP2-JPCA2023}. The key idea of OBMP2 is the use of canonical transformation\cite{CT-JCP2006,CT-JCP2007,CT-ACP2007,CT-JCP2009,CT-JCP2010,CT-IRPC2010} followed by the cumulant approximation\cite{cumulant-JCP1997,cumulant-PRA1998,cumulant-CPL1998,cumulant-JCP1999} to derive an effective one-body Hamiltonian. The resulting OBMP2 Hamiltonian is a sum of the standard Fock operator and a one-body correlation MP2 potential. Molecular orbitals and orbital energies are relaxed in the presence of correlation by diagonalizing {\it correlated} Fock matrix. The double-excitation MP2 amplitudes are then updated using those new molecular orbitals and orbital energies, resulting in a self-consistency. We have shown that the self-consistency of OBMP2 can resolve issues caused by the non-iterative nature of standard MP2 calculations for open-shell systems \cite{OBMP2-JPCA2021, OBMP2-PCCP2022}. It is also surprising that OBMP2 does not suddenly break down in bond stretching \cite{OBMP2-JPCA2023}. 

In this work, we present the extension of OBMP2 by introducing SOS into the double-excitation amplitudes, denoted as the O2BMP2 method. We assess the O2BMP2 performance on the triple-bond N$_2$ dissociation curve, singlet-triplet (ST) gaps of various sets of molecules, and ionization potentials (IPs) obtained from the Koopmans' approximation. We found that O2BMP2 can dramatically outperform standard MP2 and reach the accuracy of coupled-cluster methods in all cases considered in this work. Also, O2BMP2  performs better than OBMP2 in most cases.

Details of OBMP2 theory are presented in Refs.~\citenum{OBMP2-JPCA2021,OBMP2-PCCP2022,OBMP2-JPCA2023}, and it is implemented in a local version of PySCF\cite{pyscf-2018}. The OBMP2 Hamiltonian is derived through the canonical transformation \cite{CT-JCP2006,CT-JCP2007,CT-ACP2007,CT-JCP2009,CT-JCP2010,CT-IRPC2010}:
\begin{align}
\hat{\bar{H}} = e^{\hat{A}^\dagger} \hat{H} e^{\hat{A}},
\label{Hamiltonian:ct}
\end{align}
with the molecular Hamiltonian as
\begin{align}
  \hat{H} =  \sum_{pq}h^{p}_{q} \hat{a}_{p}^{q} + \tfrac{1}{2}\sum_{pqrs}g^{p r}_{q s}\hat{a}_{p r}^{q s}\label{eq:h1}.
\end{align}
Here, $\left\{p, q, r, \ldots \right\}$ indices referring to general ($all$) spin orbitals. One- and two-body second-quantized operators $\hat{a}_p^q$ and $\hat{a}_{pq}^{rs}$ are given by $\hat{a}_p^q = \hat{a}^\dagger_p\hat{a}_q$ and $\hat{a}_{pq}^{rs} = \hat{a}^\dagger_p\hat{a}^\dagger_q\hat{a}_s\hat{a}_r$. $h_{pq}$ and $v_{pq}^{rs}$ are one- and two-electron integrals, respectively. In OBMP2, the anti-Hermitian excited operator $\hat{A}$ includes only double excitations. 
\begin{align}
  \hat{A} = \hat{A}_\text{D} = \tfrac{1}{2} \sum_{ij}^{occ} \sum_{ab}^{vir} T_{ij}^{ab}(\hat{a}_{ab}^{ij} - \hat{a}_{ij}^{ab}) \,, \label{eq:op1}
\end{align}
with the MP2 amplitude 
\begin{align}
  T_{i j}^{a b} =  \frac{g_{i j}^{a b} } { \epsilon_{i} + \epsilon_{j} - \epsilon_{a} - \epsilon_{b} } \,, \label{eq:amp}
\end{align}
where $\left\{i, j, k, \ldots \right\}$ indices refer to occupied ($occ$) spin orbitals and
$\left\{a, b, c, \ldots \right\}$ indices refer to virtual ($vir$) spin orbitals; $\epsilon_{i}$ is the orbital energy of the spin-orbital $i$. The OBMP2 Hamiltonian is defined as

\begin{align}
  \hat{H}_\text{OBMP2} = \hat{H}_\text{HF} + \left[\hat{H},\hat{A}_\text{D}\right]_1 + \tfrac{1}{2}\left[\left[\hat{F},\hat{A}_\text{D}\right],\hat{A}_\text{D}\right]_1 = \,\, \hat{H}_\text{HF} + \hat{v}_\text{OBMP2} \label{eq:h4}.
\end{align}

In Eq.~\ref{eq:h4}, commutators with the subscription 1, $[\ldots]_1$, involve one-body operators and constants that are reduced from many-body operators using the cumulant approximation\cite{cumulant-JCP1997,cumulant-PRA1998,cumulant-CPL1998,cumulant-JCP1999}. $\hat{H}_\text{HF}$ is standard HF Hamiltonian and $\hat{v}_\text{OBMP2}$ is a correlated potential composing of one-body operators with the working expression given by
\begin{align}
\hat{v}_{\text{OBMP2}} = &  \overline{T}_{i j}^{a b} \left[ f_{a}^{i} \,\hat{\Omega}\left( \hat{a}_{j}^{b} \right) 
  + g_{a b}^{i p} \,\hat{\Omega} \left( \hat{a}_{j}^{p} \right) - g^{a q}_{i j} \,\hat{\Omega} \left( \hat{a}^{b}_{q} \right) \right] \nonumber \\  &- 2 \overline{T}_{i j}^{a b}g^{i j}_{a b} 
   + \,f_{a}^{i}\overline{T}_{i j}^{a b}\overline{T}_{j k}^{b c} \,\hat{\Omega} \left(\hat{a}_{c}^{k} \right) \nonumber \\ 
     &+  f_{c}^{a}T_{i j}^{a b}\overline{T}_{i l}^{c b} \,\hat{\Omega} \left(\hat{a}^{l}_{j} \right) + f_{c}^{a}T_{i j}^{a b}\overline{T}_{k j}^{c b} \,\hat{\Omega} \left(\hat{a}^{k}_{i} \right) \nonumber \\ 
     &-  f^{k}_{i}T_{i j}^{a b}\overline{T}_{k l}^{a b} \,\hat{\Omega} \left(\hat{a}_{l}^{j} \right)
     -  f^{p}_{i}T_{i j}^{a b}\overline{T}_{k j}^{a b} \,\hat{\Omega} \left(\hat{a}^{p}_{k} \right) \nonumber \\ 
     & +  f^{k}_{i} T_{i j}^{a b}\overline{T}_{k j}^{a d} \,\hat{\Omega}\left(\hat{a}_{b}^{d} \right) +  f_{k}^{i}T_{i j}^{a b}\overline{T}_{k j}^{c b} \,\hat{\Omega} \left(\hat{a}_{a}^{c} \right) \nonumber \\ 
     &-  f_{c}^{a}T_{i j}^{a b}\overline{T}_{i j}^{c d} \,\hat{\Omega} \left(\hat{a}^{b}_{d} \right) \,
     - f_{p}^{a}T_{i j}^{a b}\overline{T}_{i j}^{c b} \,\hat{\Omega} \left(\hat{a}^{p}_{c} \right) \nonumber \\
     & - 2f_{a}^{c}{T}_{i j}^{a b}\overline{T}_{i j}^{c b} +  2f_{i}^{k}{T}_{i j}^{a b}\overline{T}_{k j}^{a b}. \label{eq:vobmp2} 
\end{align}
with $\overline{T}_{ij}^{ab} = {T}_{ij}^{ab} - {T}_{ji}^{ab}$, the symmetrization operator $\hat{\Omega} \left( \hat{a}^{p}_{q} \right) = \hat{a}^{p}_{q}  + \hat{a}^{q}_{p}$, and the Fock matrix 
\begin{align}
    f_p^q = h_p^q + \sum_{i}^{occ}\left(g^{p i}_{q i} - g^{p i}_{i q} \right).
\end{align}
We rewrite $\hat{H}_\text{OBMP2}$ (Eq.\ref{eq:h4}) in a similar form to standard HF as follows:
\begin{align}
  \hat{H}_\text{OBMP2} = & \hat{\bar{F}} + \bar{C}, \label{eq:h5}
\end{align}
where the constant $\bar{C}$ is a sum of terms without excitation operators. $\hat{\bar{F}}$ is the correlated Fock operator, $\hat{\bar{F}} =  \bar{f}^{p}_{q} \hat{a}_{p}^{q}$, with
correlated Fock matrix $\bar{f}^{p}_{q}$ written as
\begin{align}
\bar{f}^{p}_{q} &= f^{p}_{q} + v^{p}_{q}. \label{eq:corr-fock}
\end{align}
$v^{p}_{q}$ serves as the correlation potential altering the uncorrelated HF picture. The MO coefficients and energies then correspond to eigenvectors and eigenvalues of $\bar{f}^{p}_{q}$.

Grimme\cite{SCSMP2-2003} found that the MP2 performance can be dramatically improved by separating and scaling 
same-spin (SS) and opposite-spin (OS) contributions to the
correlation energy. Later, Jung {\it et al.} \cite{SOSMP2-JPC2004} extended Grimme's method by only considering the opposite-scaling component, SOS-MP2. Lochan and Head-Gordon \cite{O2-JCP2007} further developed the optimized second-order opposite-spin (O2) method by optimizing orbitals with the SOS-MP2 energy. Kossmann and Neese \cite{OOMP2-HFCC-2010} introduced spin-component scaling to the OO-MP2 method by scaling the SS and OS contributions to the MP2 amplitude. All these studies showed that SOS can significantly improve the performance of conventional counterparts. In the present work, we extend the OBMP2 method by incorporating the spin-opposite scaling $c_{\text{os}}$  into the double-excitation amplitude (Eq.~\ref{eq:amp})
\begin{align}
  T_{i j}^{a b} =  c_{\text{os}} \frac{g_{i j}^{a b} } { \epsilon_{i} + \epsilon_{j} - \epsilon_{a} - \epsilon_{b} } \,. \label{eq:amp_os}
\end{align}
The optimal value of $c_{\text{os}}$ for SOS-MP2 were found to be 1.3 \cite{O2-JCP2007}. In the current work, we will assess three values $c_{\text{os}} = 1.1, 1.2, \text{and } 1.3$ to find the best scaling for O2BMP2.

\begin{figure}[t!]
  \includegraphics[width=8cm]{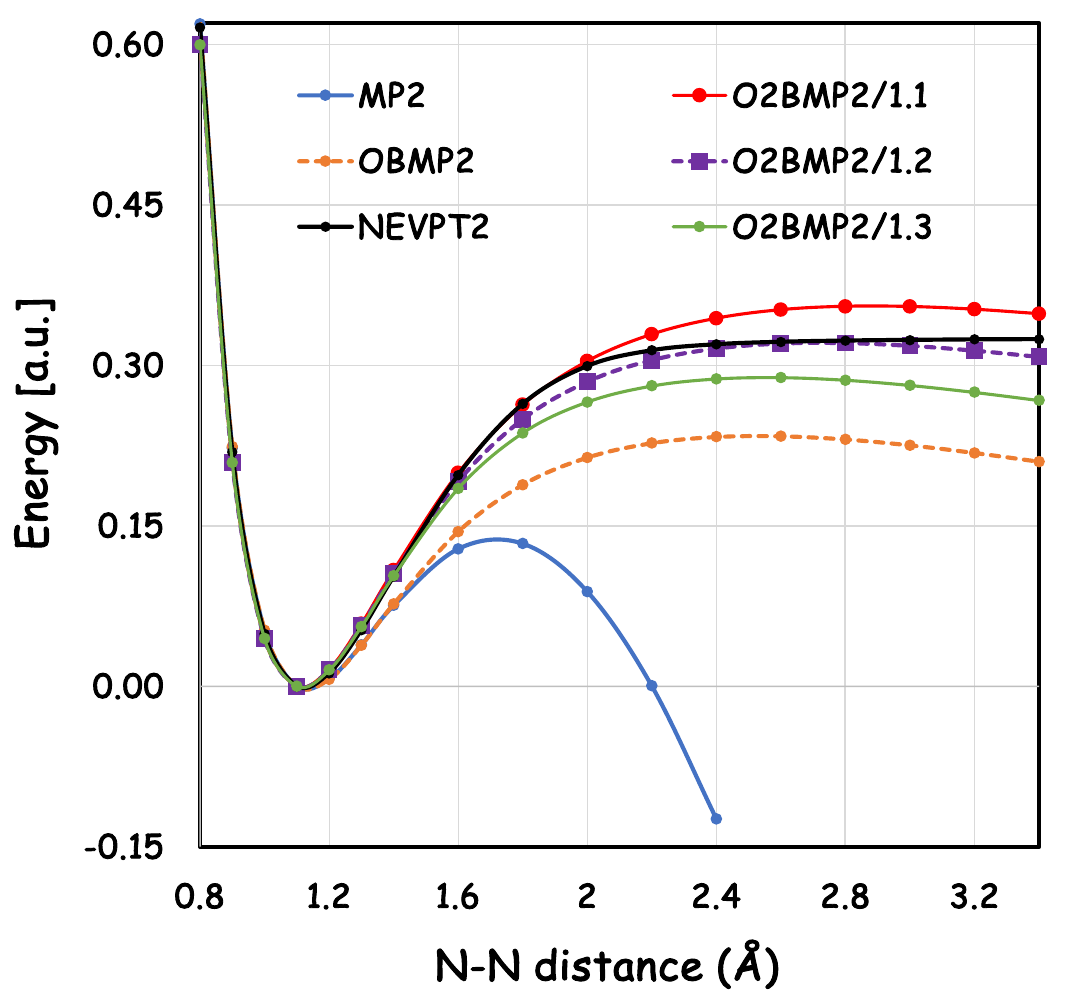}
  \caption{Potential energy curves N$_2$ molecule in cc-pVDZ from different methods. Energies are relative to the energy at the equilibrium geometry.}
  \label{fig:n2}
\end{figure}

In Ref.~\citenum{OBMP2-JPCA2023}, we showed that the self-consistency of OBMP2 helps it avoid the divergence in energy curves present in standard MP2 for H$_2$ and LiH. We now consider a more challenging system, N$_2$. We use NEVPT2 with an active space of (8e,8o) as the reference. Energies relative to the equilibrium energies of each method are presented in Figure~\ref{fig:n2}. Unsurprisingly, standard MP2 quickly breaks down, whereas OBMP2 yields a better energy curve. However, beyond the equilibrium bond length, the OBMP2 curve is far below the NEVPT2 reference. O2BMP2, with all the scaling factors considered here, can improve the energy curve upon OBMP2 and make curves close to NEVPT2. Among these factors, 1.2 may perform best, particularly at long distances.       

\begin{figure}[t!]
  \includegraphics[width=8cm]{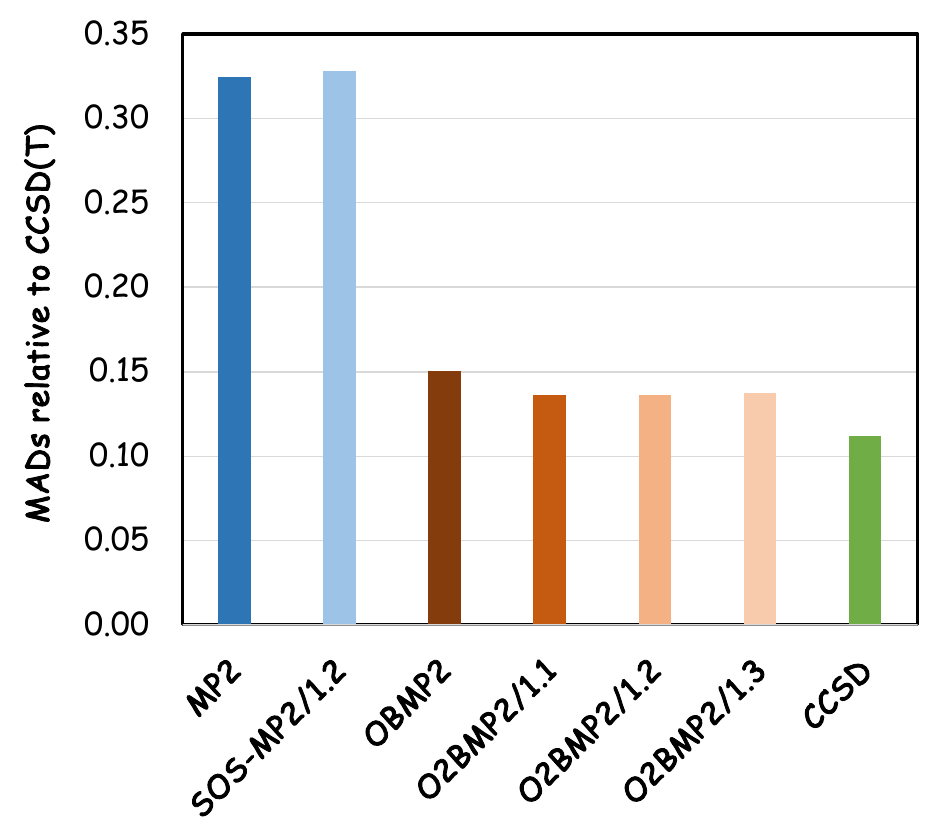}
  \caption{Mean absolute deviation (MAD) relative to CCSD(T) of singlet-triplet (ST) gaps of 38 small molecules calculated using different methods. The basis set cc-pVQZ was used.}
  \label{fig:stgap-test}
\end{figure}

\begin{figure}[h!]
  \includegraphics[width=8cm]{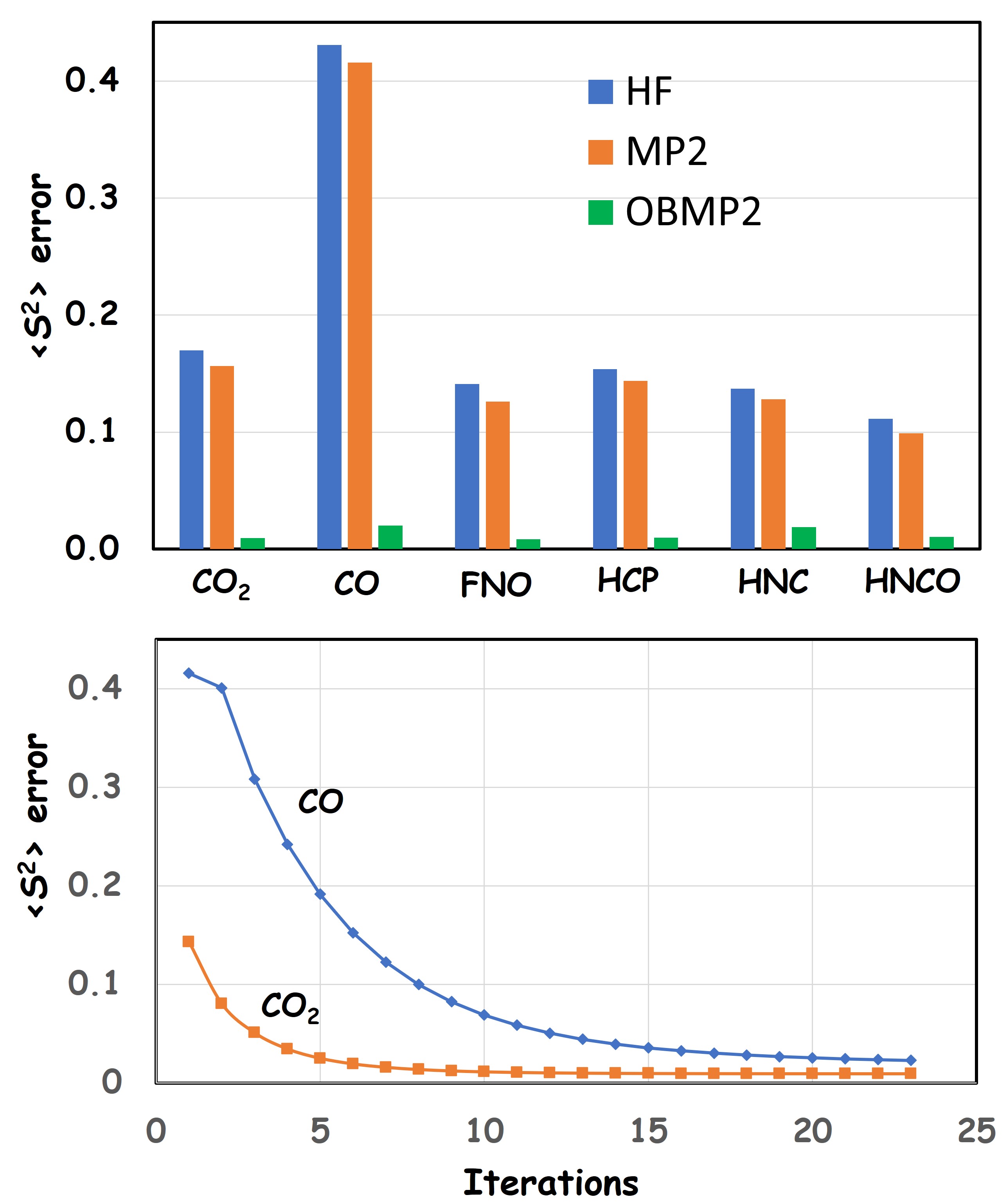}
  \caption{Upper panel: $\left<S^2\right>$ errors for the triplet state of small radicals from HF, MP2, and OBMP2. Lower panel: $\left<S^2\right>$ errors for CO and CO$_2$ at different OBMP2 iterations.}
  \label{fig:s2-test}
\end{figure}

Let us now assess the performance of our methods on the prediction of singlet-triplet (ST) gaps. We start with a test set including 38 small molecules. We first examine the spin contamination presented in Figure~\ref{fig:s2-test} in Supporting Information (SI). In the upper panel, we present some molecules for that HF severely suffers from spin contamination. We can see that while MP2 cannot eliminate the spin contamination in these cases, OBMP2 yields negligible spin contamination. In the lower panel, we plot the change in spin contamination with respect to OBMP2 iterations for two molecules CO and CO$_2$. The spin contamination at the first iteration is large and significantly reduced when the loop converges, implying the importance of self-consistency in eliminating the spin contamination. In Figure~\ref{fig:stgap-test}, we plot mean absolute deviations (MADs) relative to CCSD(T) reference of ST gaps from different methods, including MP2, SOS-MP2 with $c_\text{OS}$ = 1.2, OBMP2, and O2BMP2 with varying values of $c_\text{OS}$. ST gaps of different methods are given in SI. We can see that MP2 and SOS-MP2 give MADs larger than 0.3 eV, whereas OBMP2 and O2BMP2 with three scaling factors yield MADs smaller than 0.15 eV and comparable to CCSD. For this set of small molecules, O2BMP2 is only marginally better than OBMP2.

We now consider some medium-size organic radicals adopted from Ref.\citenum{schreiber2008benchmarks}. All results are shown in Table~\ref{tab:STgap-organic}. We compare our results to experimental values and other calculated results, including MP2, SOS-MP2($c_\text{OS}$ = 1.2), CC2, and CCSD. MP2 and SOS-MP2 yield significant errors relative to the experiment. While OBMP2 can dramatically improve MP2 ST gaps, its errors are still quite large. Interestingly, O2BMP2 with $c_{\text{os}} = 1.2$ performs better than OBMP2 with a smaller MAD (0.19 eV).   

\begin{table*}[t!]
  \normalsize
  \caption{\label{tab:STgap-organic} \normalsize Singlet-triplet gaps (in eV) of biradicals. CCSD, CC2, and experimental ST gaps are taken from Ref.~\citenum{schreiber2008benchmarks}. The basis set cc-pVTZ was employed.}
  \begin{tabular}{ccccccccccccccccccccc}
    \hline \hline		
    Molecules &exp	&CCSD	&CC2	&MP2	&SOS-MP2/1.2	&OBMP2	&O2BMP2/1.2 \\
    \hline
ethene	        &4.36	&4.42	&4.52	&4.59	&4.55	&4.60	&4.35 \\
butadiene	    &3.22	&3.25	&3.34	&3.52	&3.55	&3.44	&3.37 \\
hexatriene	    &2.61	&2.62	&2.78	&3.54	&3.51	&2.86	&2.82 \\
octatetraene	&2.1	&2.23	&2.4	&3.07	&3.06	&2.46	&2.45 \\
cyclopropene	&4.16	&4.3	&4.44	&4.52	&4.49	&4.45	&4.24 \\
cyclopentadiene	&3.1	&3.18	&3.36	&3.51	&3.46	&3.41	&3.26 \\
furan	        &4.02	&4.17	&4.30	&4.51	&4.33	&4.42	&4.11 \\
pyrrole	        &4.21	&4.52	&4.68	&4.88	&4.66	&4.76	&4.44 \\
tetrazine	    &1.69	&1.99	&1.86	&2.10	&2.63	&1.52	&2.10 \\
\hline
MAD		&&0.11		&0.25	&0.53	&0.53	&0.31	&0.19 \\
MAX		&&0.31		&0.47	&0.97	&0.96	&0.55	&0.41 \\
\hline \hline
\end{tabular}
\end{table*}

\begin{figure}[h!]
  \includegraphics[width=10cm]{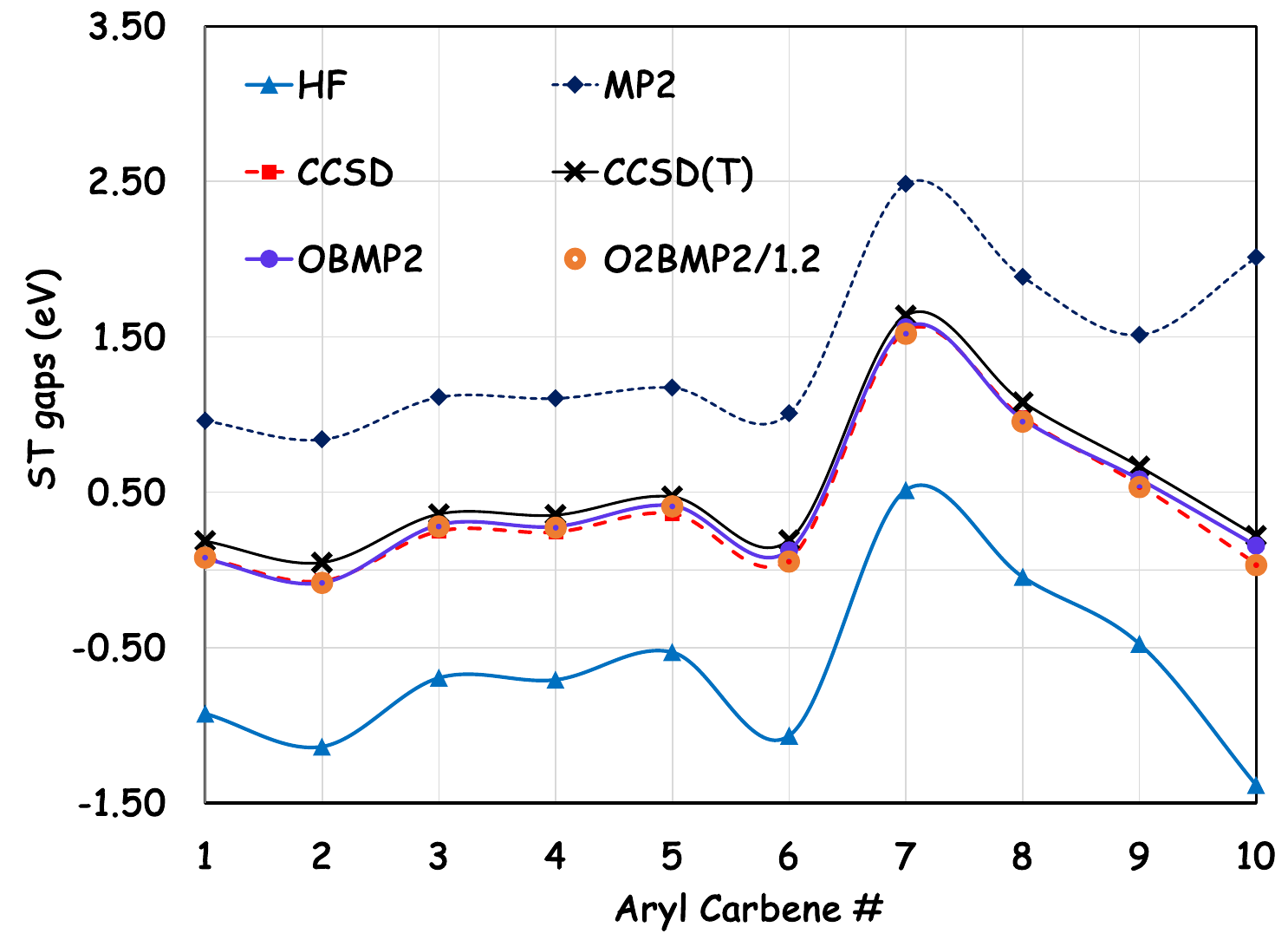}
  \caption{ST gaps of 10 aryl carbenes from different methods. The basis set cc-pVDZ was used.}
  \label{fig:stgap-aryl}
\end{figure}

\begin{figure}[t!]
  \includegraphics[width=8cm]{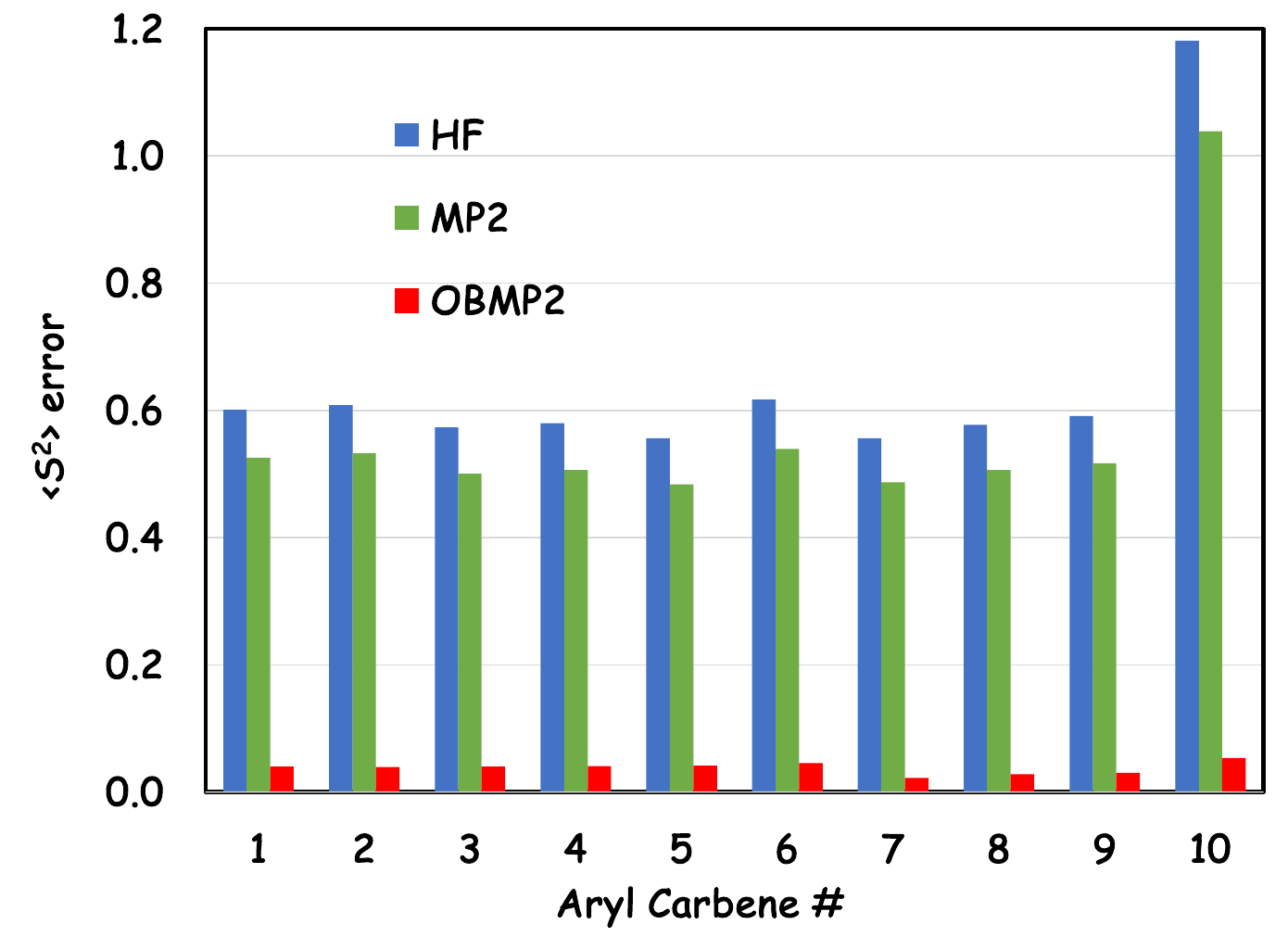}
  \caption{$\left<S^2\right>$ errors for the triplet state of 10 aryl carbenes from HF, MP2, and OBMP2}
  \label{fig:s2-aryl}
\end{figure}

\begin{figure}[t!]
  \includegraphics[width=8cm]{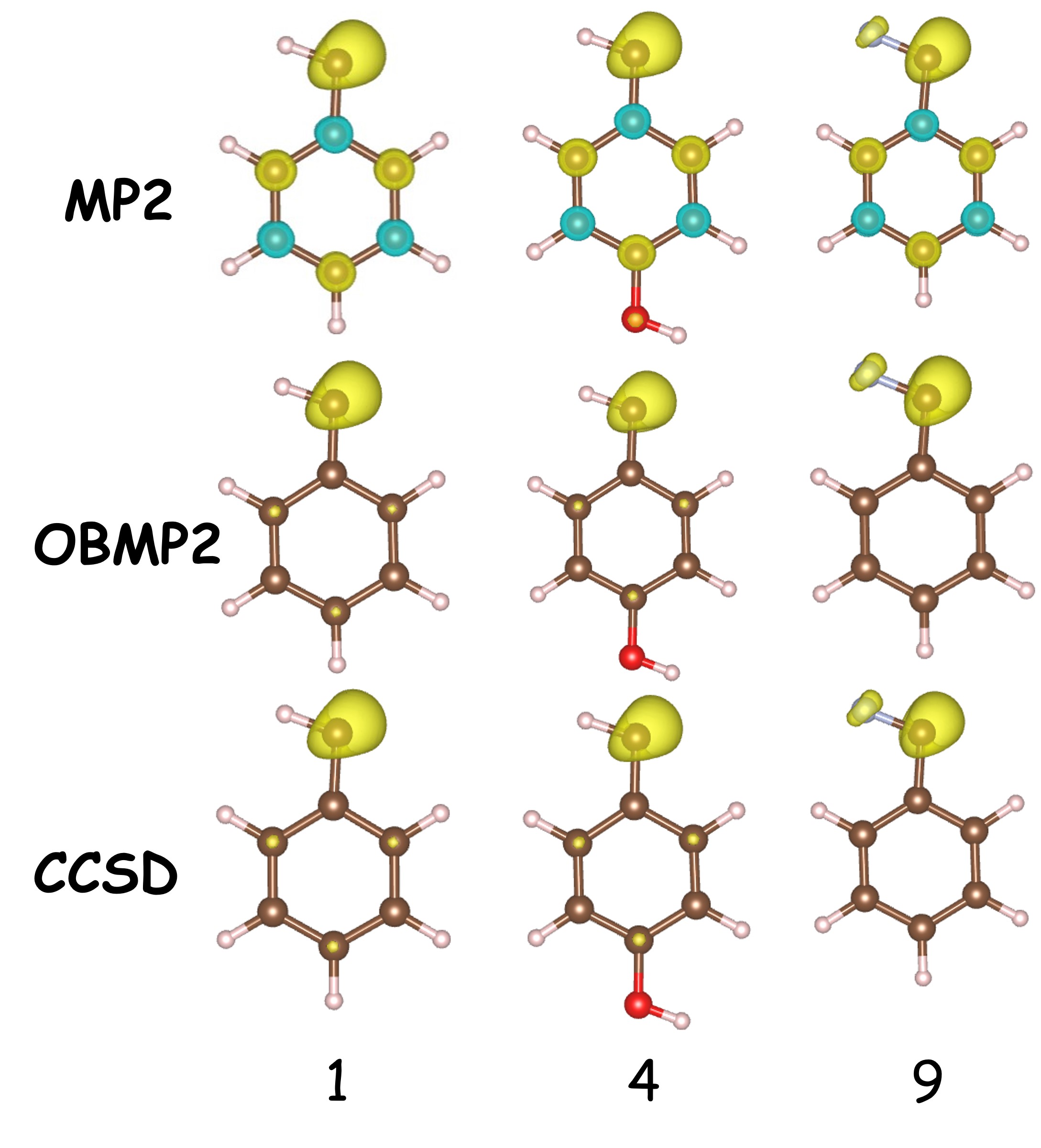}
  \caption{Spin densities of the triplet state of three aryl carbenes 1, 4, and 9 from MP2, OBMP2, and CCSD.}
  \label{fig:spindens-aryl}
\end{figure}

The next set consists of 10  aryl carbenes adopted from Ref.~\citenum{ghafarian2018accurate}. Determining the ST gap of carbenes is a difficult task for both experiment and theory. Among classes of carbenes, aryl carbenes have attracted extensive attention due to the accessibility of the triplet state. It has been evident that HF theory fails to accurately reproduce ST gaps of carbenes, whereas DFT cannot guarantee consistently accurate predictions. One of the reasons for the failure of HF and DFT in the ST gap prediction of carbenes may be the large spin contamination. As shown in Figure~\ref{fig:s2-aryl}, both HF and MP2 severely suffer from spin contamination. Thanks to the self-consistency, OBMP2 can significantly reduce the spin contamination. To further see the importance of self-consistency, we plot in Figure~\ref{fig:spindens-aryl} spin densities of three aryl carbenes 1, 4, and 9. We can see that MP2 predicts spin densities spreading over whole molecules, which may lead to large spin contamination. On the other hand, OBMP2 predicts spin densities localizing on the aryl group, which is consistent with CCSD prediction. The ST gaps of aryl carbenes predicted by MP2, OBMP2, CCSD, and CCSD(T) are presented in Figure~\ref{fig:stgap-aryl}. We use the scaling $c_{\text{os} = 1.2}$ for O2BMP2. Unsurprisingly, CCSD results are close to the CCSD(T) reference. On the other hand, while HF underestimates the ST gaps, MP2 significantly overestimates them. Our methods yield results very close to higher-cost methods, CCSD and CCSD(T). 

\begin{figure*}[t!]
  \includegraphics[width=12cm]{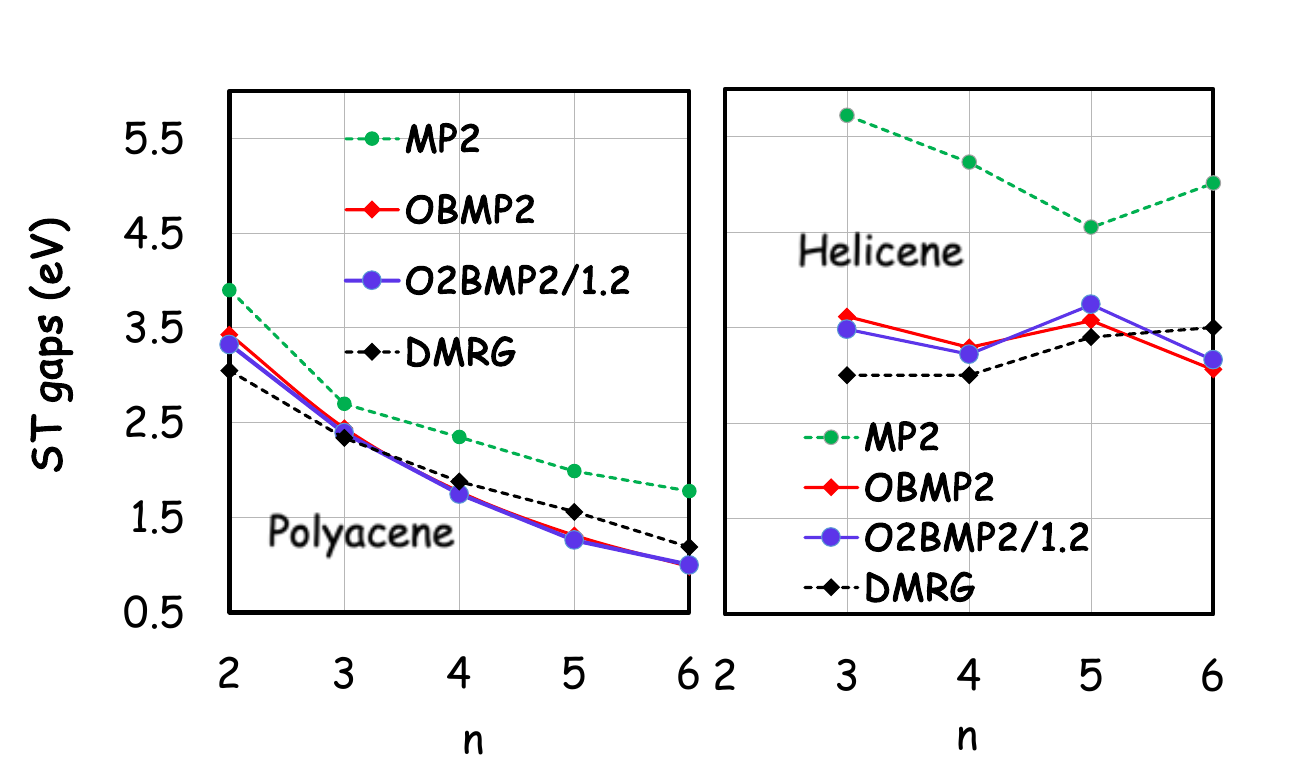}
  \caption{ST gaps of linear polyacenes (left) and helicene (right) from different methods. The DMRG references are taken from Ref.~\citenum{dey2022curious}. The basis set cc-pVDZ was used.}
  \label{fig:stgap-pahs}
\end{figure*}

\begin{figure}[t!]
  \includegraphics[width=10cm]{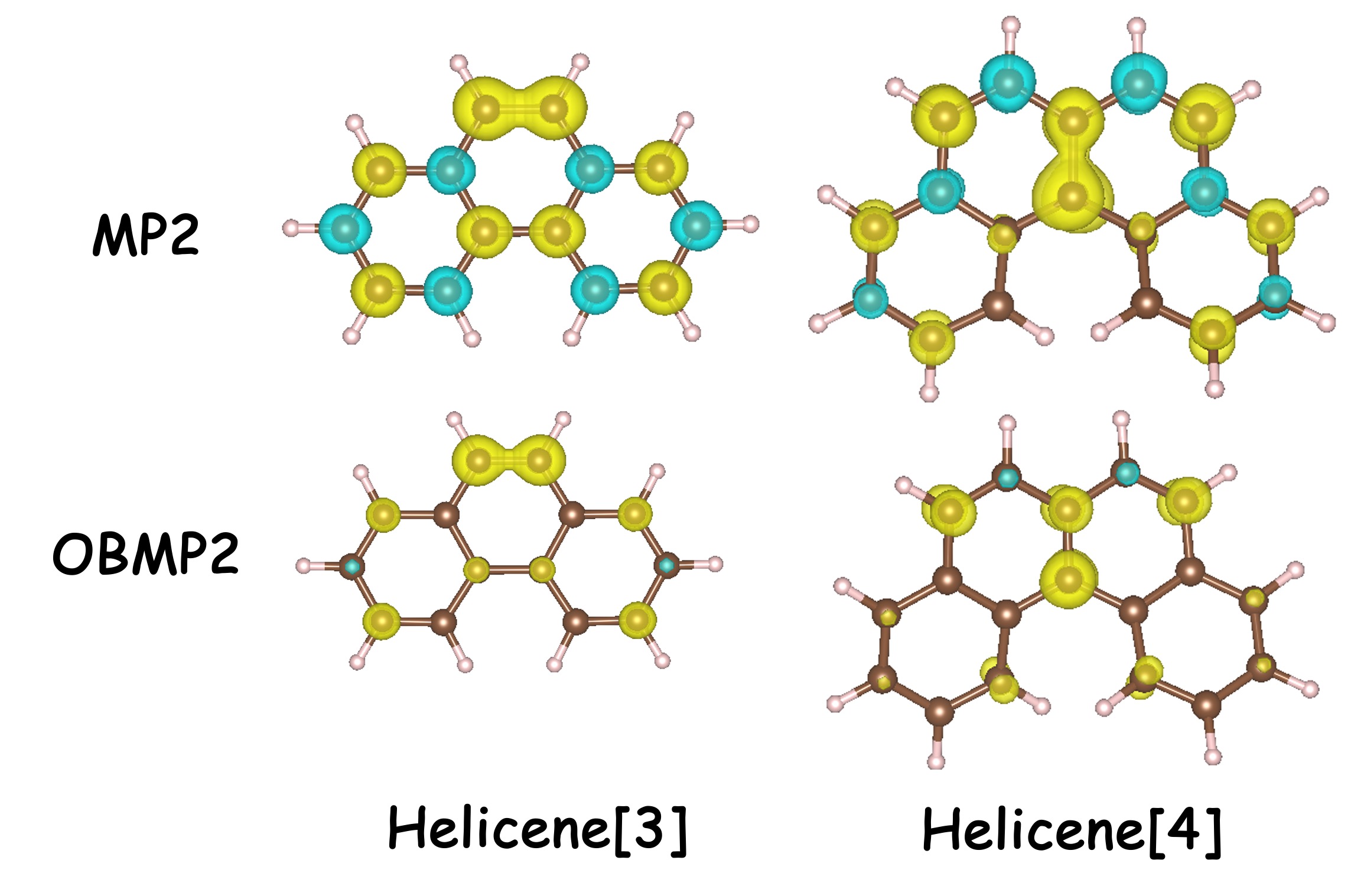}
  \caption{Spin densities of the triplet state of helicene[3] and helicene[4] from MP2 and OBMP2.}
  \label{fig:spindens-pahs}
\end{figure}

The last set we used to test the OBMP2 and O2BMP2 prediction of ST gap is polyaromatic hydrocarbons (PAHs). The prediction of accurate ST gaps of polyaromatic hydrocarbons has been challenging for computational methods\cite{hachmann2007radical, ibeji2015singlet, sharma2019density, shee2019singlet, dey2022curious}. While the ST gaps of linear PAHs have shown an exponential decay with system size, those of the non-linear PAHs are marginally sensitive to system size\cite{dey2022curious}. Unfortunately, the latter has not been observed by single-reference methods like DFT \cite{rulivsek2007convergence}. Dey and Gosh have attributed the failure of DFT to the multi-reference nature of each state of non-linear PAHs \cite{dey2022curious}. In the current work, we consider polyacenes (linear PAHs) and helicene (non-linear PAHs) with geometries taken from Ref.~\citenum{dey2022curious}. ST gaps from OBMP2 and MP2 compared to the density matrix renormalization group (DMRG) are shown in Figure~\ref{fig:stgap-pahs}. For both cases,  while MP2 errors relative to DMRG are significant, OBMP2 and O2BMP2/1.2 can dramatically improve ST gaps of PAHs. Our methods predict ST gaps close to DMRG for polyacenes, whereas their errors are still quite significant for helicene. It could be because of the stronger multi-reference nature present in helicene. It is worth stressing that OBMP2 and O2BMP2 can reproduce  DMRG prediction on the less dependence of ST gaps on the system size for helicene, which has not been observed by single-reference methods like DFT\cite{rulivsek2007convergence, dey2022curious}. In Figure~\ref{fig:spindens-pahs}, we plot spin densities of helicene[3] and helicene[4]. While MP2 spin densities are delocalized over the structures, OBMP2 ones are localized along the preferentially stable double bonds, entirely consistent with DMRG prediction \cite{dey2022curious}.  

We now move to assess the performance of our methods on the prediction of molecular IPs. Other previous studies showed that Koopmans' approximation with MP2 cannot give satisfactory accuracy in the prediction of IPs\cite{ayala2001atomic, maksic2002good}. It is interesting to check whether our methods can achieve accurate IPs in the framework of Koopmans' approximation. We previously derived the formula of OBMP2 IPs within Koopmans' approximation in Ref.~\citenum{OBMP2-JCP2013}. In the current work, we implement it in the spin-unrestricted OBMP2 version, removing only one electron instead of a pair of electrons in the restricted version.   

\begin{figure}[t!]
  \includegraphics[width=10cm]{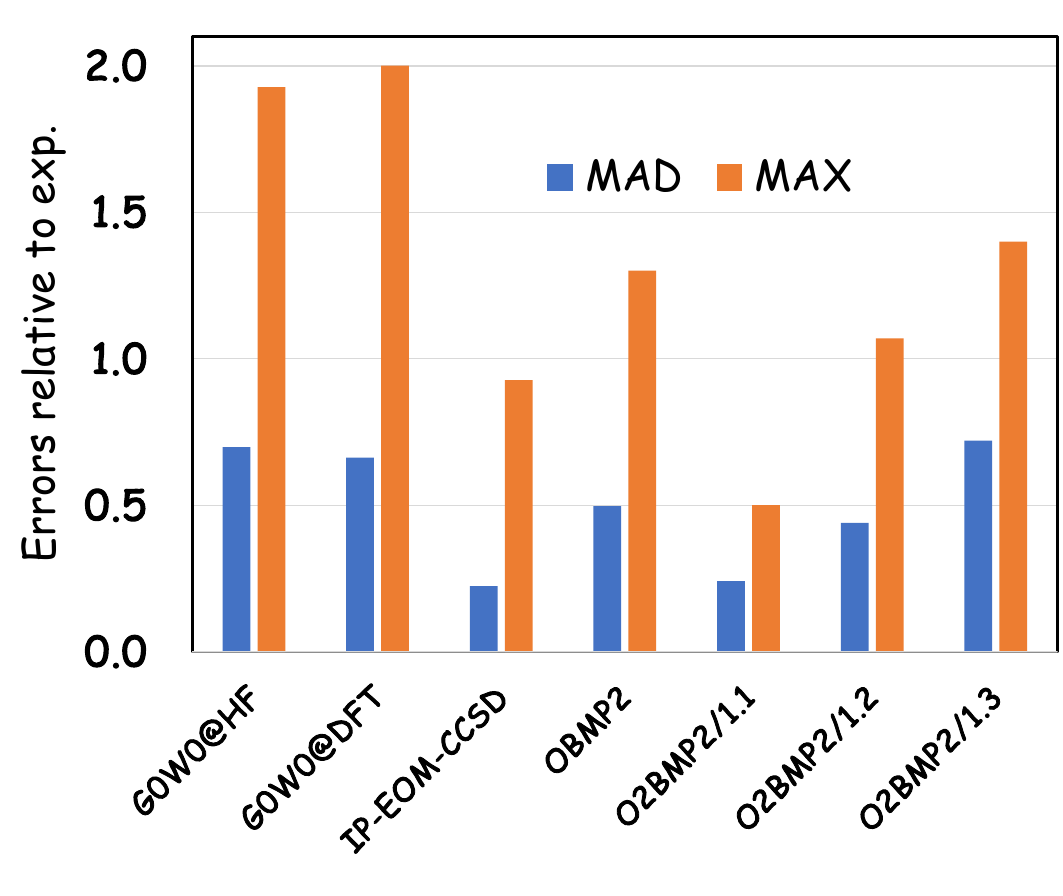}
  \caption{Mean absolute deviation (MAD) and maximum absolute deviation (MAX) relative to experimental values of valence IPs of small molecules from different methods. The basis set cc-pVQZ was used.}
  \label{fig:ips-small}
\end{figure}

We first consider a test set of 21 small molecules with 58 valence IPs. We report IP-EOM-CCSD and G0W0 with HF and DFT (PBE) references for comparison. For O2BMP2, we have tested three scaling factors $c_{\text{os}} = 1.1, 1.2, $ and $1.3$. Calculated and experimental IPs are given in SI. We show in Figure~\ref{fig:ips-small} mean absolute deviations (MAD) and maximum absolute deviations (MAX) relative to experimental values. We can see that both G0W0 yield large MAD and MAX, whereas IP-EOM-CCSD can significantly reduce MAD to 0.23 eV. Although OBMP2 performs better than G0W0, its errors are still large. Regarding O2BMP2, unlike ST gaps,  IPs are sensitive to the scaling factor $c_{\text{os}}$. Among the three values, 1.1 gives the smallest errors with MDA comparable to that of IP-EOM-CCSD and MAX even smaller than IP-EOM-CCSD.

We finally evaluate the IPs of 10 organic acceptor molecules with medium size adopted from  Ref.~\citenum{knight2016accurate}. The above assessment for small molecules shows that O2BMP2/1.1 performs best. We thus report its results in comparison with IP-EOM-CCSD and G0W0@HF. All results are summarized in Table~\ref{tab:IP-organic}. G0W0@HF vastly overestimates IPs of acceptor molecules with MAD up to 0.46 eV, consistent with the error found in Ref.~\citenum{knight2016accurate}. IP-EOM-CCSD yields results close to experimental values with MAD of 0.17 eV. Surprisingly, O2BMP2/1.1 can reach an accuracy similar to EOM-CCSD with MAD of 0.16 eV. The maximum error of O2BMP2/1.1 is 0.3 eV for mDCNB and benzonitrile molecules that may have a strong multi-reference nature. 

\begin{table*}[t!]
  \normalsize
  \caption{\label{tab:IP-organic} \normalsize First ionization potential of 10 organic acceptor molecules. Experimental values are adopted from Ref.~\citenum{knight2016accurate}. The basis set aug-cc-pVDZ was employed.}
  \begin{tabular}{ccccccccccccccccccccc}
    \hline \hline		
    Molecules &exp	&IP-EOM-CCSD	&G0W0@HF	&O2BMP2/1.1	\\
    \hline
Bezonquinone (BQ)	&9.95	&10.04	&10.30	&10.10 \\
tetrafluoro-BQ	&10.70	&11.05	&11.40	&10.84 \\
tetrachloro-BQ	&9.74	&10.10	&10.40	&9.88 \\
fumaronitrie	&11.15	&11.30	&11.38	&11.10 \\
maleic anhydride	&11.07	&11.02	&12.15	&10.96 \\
mDCNB	&10.20	&10.26	&10.77	&9.90 \\
nitrobenzene	&9.86	&10.06	&10.10	&9.68 \\
phtalic anhydride	&10.10	&10.40	&10.50	&10.03 \\
TCNE	&11.77	&11.91	&12.03	&11.65 \\
benzonitrile	&9.70	&9.73	&9.78	&9.40 \\
\hline
MAD		&&0.17	&0.46	&0.16 \\
MAX		&&0.36	&1.08	&0.30 \\
\hline \hline
\end{tabular}
\end{table*}

In summary, we have extended our recently developed method, OBMP2, by introducing the spin-opposite scaling to the double-excitation amplitudes, termed O2BMP2. We assess the O2BMP2 performance on the triple-bond N$_2$ dissociation, ST gaps, and IPs of medium-size organic compounds. O2BMP2 performs much better than standard MP2 and reaches the accuracy of coupled-cluster methods in all cases considered in this work. Our method is then expected to help tackle realistic, challenging systems with large sizes. Working on further reducing computational costs of OBMP2 and O2BMP2 is in progress.

\bibliography{main}
\clearpage
\beginsupplement
\section*{Supporting Information to: \\ 
Reaching high accuracy for energetic properties at second-order perturbation cost by merging self-consistency and spin-opposite scaling}

{\bf Nhan Tri Tran} \\
{\it University of Science, Vietnam National University, Ho Chi Minh City, Vietnam}\\
{\bf Hoang Thanh Nguyen} \\
{\it Institute of Applied Mechanics and Informatics, Vietnam Academy of Science and Technology, Ho Chi Minh City, Vietnam}\\
{\bf Lan Nguyen Tran*} \\
{\it Email: tnlan@hcmiu.edu.vn \\
Department of Physics, International University, Ho Chi Minh City, Vietnam \\
Vietnam National University, Ho Chi Minh City, Vietnam}

File supporting-information.xls includes:
\begin{itemize}
    \item Singlet-triplet gaps (in eV) of 39 small molecules
    \item Valence IPs (in eV) of 21 small molecules.
\end{itemize}

\clearpage

\end{document}